\documentclass[12pt]{article}
\usepackage{epsfig}
\begin{document}
%%%%%%%%%%%%%%%%%%%%%%%%%%%%%%%%%%%%%%%%%%%%%%%%%%%%%%%%%%%%%%%%%%%%%%%%%
\title{Near-threshold $K^*(892)^+$ meson production in the interaction of $\pi^-$ mesons with nuclei}
\author{E. Ya. Paryev$^{1,2}$\\
{\it $^1$Institute for Nuclear Research, Russian Academy of Sciences,}\\
{\it Moscow 117312, Russia}\\
{\it $^2$Institute for Theoretical and Experimental Physics,}\\
{\it Moscow 117218, Russia}}
%%==============================================================
%%==============================================================

\renewcommand{\today}{}
\maketitle

\begin{abstract}
We study the inclusive strange vector meson $K^*(892)^+$ production in ${\pi^-}A$ reactions at near-threshold
laboratory incident pion momenta of 1.4--2.0 GeV/c within a nuclear spectral function approach.
The approach accounts for incoherent primary $\pi^-$ meson--proton
${\pi^-}p \to {K^*(892)^+}\Sigma^-$ production processes as well as the influence of the scalar
$K^*(892)^+$--nucleus potential (or the $K^*(892)^+$ in-medium mass shift) on these processes.
We calculate the absolute differential and total cross
sections for the production of $K^*(892)^+$ mesons off carbon and tungsten nuclei at laboratory angles of
0$^{\circ}$--45$^{\circ}$ and at these momenta
within five scenarios for the above shift. We show that the $K^*(892)^+$ momentum distributions and their
excitation functions (absolute and relative) possess a high sensitivity to changes in the in-medium
$K^*(892)^+$ mass shift in the low-momentum region of 0.1--0.6 GeV/c.
Therefore, the measurement of such observables in a dedicated experiment at
the GSI pion beam facility in the near-threshold momentum domain will allow to get valuable information
on the $K^*(892)^+$ in-medium properties.
\end{abstract}

\newpage

\section*{1 Introduction}

\hspace{0.5cm} The study of the modification of the hadronic properties (masses and widths) of
light non-strange vector mesons $\rho$, $\omega$, $\phi$, light strange pseudoscalar mesons $K$ and ${\bar K}$,
pseudoscalar mesons $\eta$, $\eta^{\prime}$ as well as mesons with open and hidden charm $D$ and $J/\psi$
in a strongly interacting environment has received considerable interest in recent years owing to the
expectation to observe a partial restoration of chiral symmetry in a nuclear medium
(see, for example, [1--13]). The in-medium properties of hyperons at finite density have also been matter
of intense theoretical investigations in the last two decades [14--24]. Another interesting case of medium
renormalization of hadrons is that of the strange vector $K^*(892)$ and axial-vector $K_1(1270)$ mesons
with the same charge states (or with the same quark structure $q{\bar s}$ or ${\bar q}s$ with $q=u,d$),
whose in-medium mass difference , as is expected [25--27], is sensitive to the chiral order parameter and,
hence, will give the possibility to identify unambiguously the effect of chiral symmetry breaking in nuclear medium.
The $K^*(892)$ and the $K_1(1270)$ mesons in the quark model are a kaonic excitations with angular momenta one
and with opposite parities. Namely, their isospins, spins-parities are $I(J^P)=\frac{1}{2}(1^-)$ for the
$K^*(892)$ and $I(J^P)=\frac{1}{2}(1^+)$ for the $K_1(1270)$. They are chiral partners and have relatively
large vacuum decay widths of 50 and 90 MeV, respectively, corresponding to a mean live-times of 4 and 2.2 fm/c.

 On the theoretical side, in literature there are a lot of publications devoted to the study of the in-medium
properties  of hadronic resonances $K^*(892)$ and $K_1(1270)$. Thus, the properties of ${\bar K}^*(892)$ and
$K^*(892)$ mesons in cold nuclear matter have been investigated in Refs. [28--30] and [30, 31], respectively,
on the basis of chirally motivated model of the meson selfenergies. In particular, it was shown that the
${\bar K}^*(892)$ in-medium width is enlarged beyond 200 MeV at normal nuclear matter density $\rho_0$,
whereas that of $K^*(892)$ is barely influenced by nuclear matter. The model predicts also for the
${\bar K}^*(892)$ and $K^*(892)$ mesons, respectively, a moderately attractive and repulsive real
low-energy nuclear potentials (or their in-medium mass shifts) of about -50 and +40 MeV at density $\rho_0$.
These are similar to those for light strange mesons ${\bar K}$ and $K$. On the other hand in contrast, a negative mass
shift of about -20 MeV has been predicted for the $K^*(892)^+$
meson, having the same quark composition $u{\bar s}$ as the $K^+$ one, at rest at saturation density $\rho_0$
within the quark-meson coupling model [32]. Mass shifts of the $K^*(892)$ and $K_1(1270)$ mesons of about -40
and -150 MeV at density $\rho_0$ were obtained in a more recent calculations in the framework of
the three-flavor extended linear sigma model [33]. Other recent calculations performed in Ref. [26] using QCD
sum rule show that the upper limits of the mass shifts of $K_1^-$ and $K_1^+$ mesons in nuclear matter are
-249 and -35 MeV, respectively.

What concerns the experimental situation, up to now only a scarce data on $K^*$ production in heavy-ion
and proton--proton collisions have been collected in the experiments performed in the SPS [34], RHIC [35]
and LHC [36] energy domains. At SIS energies, the subthreshold and deep subthreshold production of
$K^*(892)^0$ mesons in Al+Al and Ar+KCl collisions at a beam kinetic energies of 1.9 A GeV and 1.76 A GeV,
respectively, has been reported by the FOPI [37] and the HADES [38] Collaborations. While the
$K^*(892)^0$/$K^0$ yield ratio deduced in the FOPI experiment is found in good agreement with the corresponding
prediction of the UrQMD transport model, this ratio extracted in the HADES experiment is overestimated by the model
by factor of about two. Probably, less discrepancy might appear here if in-medium modifications of kaon properties
will be implemented into this transport model.
The medium modification of the $K_1(1270)$ meson could be probed at J-PARC through the $K^-$ reaction on
various nuclear targets [26]. Such measurement together with that of $K^*(892)$ will shed light on the
partial restoration of chiral symmetry in nuclear matter [26].

As a guidance for such future dedicated experiments and as a first step in the implementation of this programme,
in the present study we give the predictions
for the absolute differential and total cross sections for near-threshold production of $K^*(892)^+$ mesons in
${\pi^-}^{12}C \to {K^*(892)^+}X$ and ${\pi^-}^{184}W \to {K^*(892)^+}X$ reactions at laboratory angles of
0$^{\circ}$--45$^{\circ}$ by
incident pions with momenta below 2.0 GeV/c as well as for their relative yields
from these reactions within different scenarios for the $K^*(892)^+$ in-medium mass shift.
These nuclear targets were employed in recent measurements [39]
of $\phi$ meson production in ${\pi^-}A$ reactions at the GSI pion beam facility using the HADES
spectrometer and, therefore, can be adopted in studying the ${\pi^-}A \to {K^*(892)^+}X$ interactions here.
The calculations are based on a first-collision model using an eikonal approximation, developed in Refs. [9, 10, 21] for the
description of the inclusive $\phi$ and $\eta^{\prime}$ meson as well as $\Lambda(1520)$ hyperon production
and extended to account for different scenarios for the $K^*(892)^+$ in-medium mass shift.
This model is based on the quasiparticle picture and, therefore, it is more appropriate for consideration
of the $K^*$ meson production in nuclei than for the study of the ${\bar K}^*$ creation here since,
contrary to the ${\bar K}^*$, the $K^*$ meson behaves in the medium as a quasiparticle with a single-peak
spectral function and a modified effective mass [30, 31].
Our calculations can be used as an important tool for possible extracting of the valuable information on the
$K^*(892)^+$ in-medium mass shift from the data which could be taken in a
dedicated experiment at the GSI pion beam facility.

\section*{2 Model: direct mechanism of $K^*(892)^+$ meson production on nuclei}

\hspace{0.5cm} Since we are interested in near-threshold incident pion beam momenta below 2.0 GeV/c,
we have accounted for the following direct elementary $K^*(892)^+$ production process which has
the lowest free production threshold momentum (1.84 GeV/c)
\footnote{$^)$ We can ignore in the momentum domain of interest the contribution to the $K^*(892)^+$
yield from the processes $\pi^-p \to {K^*(892)^+}\Lambda{\pi^-}$ and $\pi^-N \to {K^*(892)^+}\Sigma{\pi}$
due to larger their production thresholds
($\approx$ 1.97 and 2.15 GeV/c, respectively) in ${\pi^-}p$ and $\pi^-N$ collisions.
Moreover, taking into consideration the results of the study [9] of pion-induced
$\phi$ meson production on $^{12}$C and $^{184}$W nuclei at beam momentum of 1.7 GeV/c,
we neglect in this domain by analogy with [9] the secondary pion--nucleon
${\pi}N \to {K^*(892)^+}\Lambda$ and ${\pi}N \to {K^*(892)^+}\Sigma$ production processes.}$^)$
:
%formula(1)
\begin{equation}
\pi^-+p \to K^*(892)^++\Sigma^-.
\end{equation}
For numerical simplicity, in our calculations we will account for the medium modification of the final
$K^*(892)^+$ meson, participating in the production process (1), by adopting its average in-medium mass
$<m_{K^*}^*>$ instead of its local effective mass $m_{K^*}^*(|{\bf r}|)$ in the in-medium cross section
of this process, with $<m_{K^*}^*>$ defined according to Refs. [9, 10] as:
%formula(2)
\begin{equation}
<m^*_{K^*}>=m_{K^*}+V_0\frac{<{\rho_N}>}{{\rho_0}}.
\end{equation}
Here, $m_{K^*}$ is the $K^*(892)^+$ free space mass,
$V_0$ is the $K^*(892)^+$ effective scalar
nuclear potential (or its in-medium mass shift) at normal nuclear matter
density ${\rho_0}$, and $<{\rho_N}>$ is the average nucleon density.
For target nuclei $^{12}$C and $^{184}$W,  the ratio $<{\rho_N}>/{\rho_0}$, was chosen as
0.55 and 0.76, respectively, in the present work. With regards to the quantity $V_0$, we will adopt for it in line
with above-mentioned the five following options: i) $V_0=-40$ MeV, ii) $V_0=-20$ MeV, iii) $V_0=0$ MeV, iv) $V_0=+20$ MeV,
and v) $V_0=+40$ MeV throughout the study.
Following the predictions of the chiral effective field theory approach [18, 40], SU(6) quark model [41, 42] for
the fate of hyperons in nuclear matter and phenomenological information deduced from hypernuclear data [6, 43]
that the $\Sigma$ hyperon experiences only a moderately repulsive nuclear potential of about 10--40 MeV at central
nuclear densities and finite momenta as well as a weakly attractive potential at the surface of the nucleus, we
will ignore the modification of the mass of the $\Sigma^-$ hyperons, produced together with the $K^*(892)^+$
mesons in the process (1), in the nuclear medium.
Accounting for that the in-medium threshold energy
\footnote{$^)$Determining mainly the strength of the $K^*(892)^+$ production cross sections in
near-threshold pion--nucleus collisions.}$^)$
$\sqrt{s^*_{\rm th}}=<m_{K^*}^*>+m_{\Sigma^-}$
of the process (1) looks like that for the final charged particles, influenced also by the respective Coulomb
potentials, due to the cancelation of these potentials,
we will neglect here their impact on these particles as well.

The total energy $E^\prime_{K^*}$ of the $K^*(892)^+$ meson in nuclear matter is
expressed via its average effective mass $<m^*_{K^*}>$ and its in-medium momentum
${\bf p}^{\prime}_{K^*}$ by the expression [9, 10]:
%formula(3)
\begin{equation}
E^\prime_{K^*}=\sqrt{({\bf p}^{\prime}_{K^*})^2+(<m^*_{K^*}>)^2}.
\end{equation}
The momentum ${\bf p}^{\prime}_{K^*}$ is related to the vacuum $K^*(892)^+$
momentum ${\bf p}_{K^*}$ as follows [9, 10]:
%formula(4)
\begin{equation}
E^\prime_{K^*}=\sqrt{({\bf p}^{\prime}_{K^*})^2+(<m^*_{K^*}>)^2}=
\sqrt{{\bf p}^2_{K^*}+m^2_{K^*}}=E_{K^*},
\end{equation}
where $E_{K^*}$ is the $K^*(892)^+$ total energy in vacuum.

Since the $K^*(892)^+$--nucleon total cross section is expected to be small [44], we will neglect both inelastic and
quasielastic ${K^*(892)^+}N$ interactions in the present study.
Then, accounting for the distortion of the incident pion in nuclear matter and the attenuation of the flux
of the $K^*(892)^+$ mesons in the nucleus due to their decays here
\footnote{$^)$Eq. (9) shows that for typical values $p^{\prime}_{K^*} \approx m^*_{K^*}$ and vacuum total
$K^*(892)^+$ decay width in its rest frame $\Gamma_{K^*}=50$ MeV the $K^*(892)^+$ decay mean free path $\lambda_{K^*}$
is equal to 4 fm. This value is comparable with the radius of $^{12}$C of 3 fm and it is much less than that
of $^{184}$W of 7.4 fm.}$^)$
as well as using the results given in [9, 10, 21], we represent the inclusive differential
cross section for the production of $K^*(892)^+$ mesons with vacuum momentum
${\bf p}_{K^*}$ on nuclei in the direct process (1) as follows:
%formula(5)
\begin{equation}
\frac{d\sigma_{{\pi^-}A \to {K^*(892)^+}X}^{({\rm prim})}
({\bf p}_{\pi^-},{\bf p}_{K^*})}
{d{\bf p}_{K^*}}=I_{V}[A,\theta_{K^*}]
\left(\frac{Z}{A}\right)\left<\frac{d\sigma_{{\pi^-}p\to K^*(892)^+{{\Sigma^-}}}({\bf p}_{\pi^-},
{\bf p}^{\prime}_{{K^*}})}{d{\bf p}^{\prime}_{{K^*}}}\right>_A\frac{d{\bf p}^{\prime}_{{K^*}}}
{d{\bf p}_{{K^*}}},
\end{equation}
where
%formula(6)
\begin{equation}
I_{V}[A,\theta_{{K^*}}]=A\int\limits_{0}^{R}r_{\bot}dr_{\bot}
\int\limits_{-\sqrt{R^2-r_{\bot}^2}}^{\sqrt{R^2-r_{\bot}^2}}dz
\rho(\sqrt{r_{\bot}^2+z^2})
\exp{\left[-\sigma_{{\pi^-}N}^{\rm tot}A\int\limits_{-\sqrt{R^2-r_{\bot}^2}}^{z}
\rho(\sqrt{r_{\bot}^2+x^2})dx\right]}
\end{equation}
$$
\times
\int\limits_{0}^{2\pi}d{\varphi}\exp{\left[-
\int\limits_{0}^{l(\theta_{{K^*}},\varphi)}\frac{dx}
{\lambda_{K^*}(\sqrt{x^2+2a(\theta_{{K^*}},\varphi)x+b+R^2})}\right]},
$$
%formula(7)
\begin{equation}
a(\theta_{K^*},\varphi)=z\cos{\theta_{K^*}}+
r_{\bot}\sin{\theta_{K^*}}\cos{\varphi},\\\
b=r_{\bot}^2+z^2-R^2,
\end{equation}
%formula(8)
\begin{equation}
l(\theta_{K^*},\varphi)=\sqrt{a^2(\theta_{K^*},\varphi)-b}-
a(\theta_{K^*},\varphi),
\end{equation}
%formula(9)
\begin{equation}
\lambda_{K^*}(|{\bf r}|)=\frac{p^{\prime}_{K^*}}{m^*_{K^*}(|{\bf r}|)\Gamma_{K^*}},\,\,\,
m^*_{K^*}(|{\bf r}|)=m_{K^*}+V_0\frac{{\rho_N}(|{\bf r}|)}{{\rho_0}}
\end{equation}
and
%formula(10), before (7)
\begin{equation}
\left<\frac{d\sigma_{{\pi^-}p\to K^*(892)^+{\Sigma^-}}({\bf p}_{\pi^-},{\bf p}^{\prime}_{K^*})}
{d{\bf p}^{\prime}_{K^*}}\right>_A=
\int\int
P_A({\bf p}_t,E)d{\bf p}_tdE
\end{equation}
$$
\times
\left\{\frac{d\sigma_{{\pi^-}p\to K^*(892)^+{\Sigma^-}}[\sqrt{s},<m_{K^*}^*>,
m_{\Sigma^-},{\bf p}^{\prime}_{K^*}]}
{d{\bf p}^{\prime}_{K^*}}\right\},
$$
%formula(11), before (8)
\begin{equation}
  s=(E_{\pi^-}+E_t)^2-({\bf p}_{\pi^-}+{\bf p}_t)^2,
\end{equation}
%formula(12), before (9)
\begin{equation}
   E_t=M_A-\sqrt{(-{\bf p}_t)^2+(M_{A}-m_{N}+E)^{2}}.
\end{equation}
Here,
$d\sigma_{{\pi^-}p\to {K^*(892)^+}{\Sigma^-}}[\sqrt{s},<m_{K^*}^*>,m_{\Sigma^-},{\bf p}^{\prime}_{K^*}]
/d{\bf p}^{\prime}_{K^*}$
is the off-shell inclusive differential cross section for the production of ${K^*(892)^+}$ meson and
$\Sigma^-$ hyperon with modified mass $<m_{K^*}^*>$ and free mass $m_{\Sigma^-}$, respectively.
The $K^*(892)^+$ meson is produced with in-medium momentum
${\bf p}^{\prime}_{{K^*}}$ in process (1) at the ${\pi^-}p$ center-of-mass energy $\sqrt{s}$.
$E_{\pi^-}$ and ${\bf p}_{\pi^-}$ are the total energy and momentum of the incident pion
($E_{\pi^-}=\sqrt{m^2_{\pi}+{\bf p}^2_{\pi^-}}$, $m_{\pi}$ is the free space pion mass);
$\rho({\bf r})$ and $P_A({\bf p}_t,E)$ are the local nucleon density and the
spectral function of the target nucleus A normalized to unity
(the concrete information about these quantities, used in the subsequent calculations, is given
in Refs. [9, 45--47]);
${\bf p}_t$ and $E$ are the internal momentum and removal energy of the struck target proton
involved in the collision process (1); $\sigma_{{\pi^-}N}^{\rm tot}$ is the total cross section of the
free ${\pi^-}N$ interaction (we use in our calculations the value of $\sigma_{{\pi^-}N}^{\rm tot}=35$ mb
for initial pion momenta of interest); $Z$ and $A$ are the numbers of protons and nucleons in
the target nucleus, and $M_{A}$  and $R$ are its mass and radius; $m_N$ is the free space nucleon mass;
and $\theta_{K^*}$ is the polar angle of
vacuum momentum ${\bf p}_{{K^*}}$ in the laboratory system with z-axis directed along the momentum
${\bf p}_{{\pi^-}}$ of the incident pion beam.

   In line with [9], we assume that the off-shell differential cross section
$d\sigma_{{\pi^-}p \to {K^*(892)^+}{\Sigma^-}}[\sqrt{s},<m_{K^*}^*>,m_{\Sigma^-},{\bf p}^{\prime}_{K^*}]
/d{\bf p}^{\prime}_{K^*}$ for $K^*(892)^+$
production in channel (1) is equivalent to
the respective on-shell cross section calculated for the off-shell kinematics of this channel
as well as for the final $K^*(892)^+$ and hyperon in-medium mass $<m_{K^*}^*>$ and free mass $m_{\Sigma^-}$,
respectively. Accounting for the two-body kinematics of the process (1),
we obtain the following expression for the differential cross section
$d\sigma_{{\pi^-}p \to {K^*(892)^+}{\Sigma^-}}[\sqrt{s},<m_{K^*}^*>,m_{\Sigma^-},{\bf p}^{\prime}_{K^*}]
/d{\bf p}^{\prime}_{K^*}$:
%FORMULA (13), before (10)
\begin{equation}
\frac{d\sigma_{\pi^{-}p \to {K^*(892)^+}{\Sigma^-}}[\sqrt{s},<m_{K^*}^*>,m_{\Sigma^-},{\bf p}^{\prime}_{K^*}]}
{d{\bf p}^{\prime}_{K^*}}=
\frac{\pi}{I_2[s,<m_{K^*}^*>,m_{\Sigma^-}]E^{\prime}_{K^*}}
\end{equation}
$$
\times
\frac{d\sigma_{{\pi^{-}}p \to {K^*(892)^+}{\Sigma^-}}(\sqrt{s},<m_{K^*}^*>,m_{\Sigma^-},\theta^*_{K^*})}
{d{\bf \Omega}^*_{K^*}}
$$
$$
\times
\frac{1}{(\omega+E_t)}\delta\left[\omega+E_t-\sqrt{m_{\Sigma^-}^2+({\bf Q}+{\bf p}_t)^2}\right],
$$
where
%FORMULA (14), before (11)
\begin{equation}
I_2[s,<m_{K^*}^*>,m_{\Sigma^-}]=\frac{\pi}{2}
\frac{\lambda[s,(<m_{K^*}^*>)^{2},m_{\Sigma^-}^{2}]}{s},
\end{equation}
%FORMULA (15), before (12)
\begin{equation}
\lambda(x,y,z)=\sqrt{{\left[x-({\sqrt{y}}+{\sqrt{z}})^2\right]}{\left[x-
({\sqrt{y}}-{\sqrt{z}})^2\right]}},
\end{equation}
%FORMULA (16), before (13)
\begin{equation}
\omega=E_{\pi^-}-E^{\prime}_{K^*}, \,\,\,\,{\bf Q}={\bf p}_{\pi^-}-{\bf p}^{\prime}_{K^*}.
\end{equation}
Here,
$d\sigma_{{\pi^{-}}p \to {K^*(892)^+}{\Sigma^-}}(\sqrt{s},<m_{K^*}^*>,m_{\Sigma^-},\theta^*_{K^*})
/d{\bf \Omega}^*_{K^*}$
is the off-shell differential cross section for the production of $K^*(892)^+$ mesons
in process (1) under the polar angle $\theta^*_{K^*}$ in the ${\pi^-}p$ c.m.s. It
is assumed to be isotropic in our calculations of $K^*(892)^+$ meson production in ${\pi^-}A$ reactions
\footnote{$^)$It should be pointed out that the use in the calculations for the in-medium
$K^*(892)^+$ angular distribution of the same anisotropic form as was adopted in Ref. [45] for the
${\pi^-}p \to K^+\Sigma^-$ reaction, namely:
$d\sigma_{{\pi^{-}}p \to {K^*(892)^+}{\Sigma^-}}(\sqrt{s},<m_{K^*}^*>,m_{\Sigma^-},\theta^*_{K^*})/
d{\bf \Omega}^*_{K^*}=\left[1+|\cos{\theta^*_{K^*}}|\right]
\sigma_{{\pi^{-}}p \to {K^*(892)^+}{\Sigma^-}}(\sqrt{s},\sqrt{s^*_{\rm th}})/6\pi$
instead of isotropic one (17) leads to only insignificant corrections to the absolute $K^*(892)^+$
momentum differential cross sections presented in Figs. 2, 3 and 4. They are about 5--10\% for subthreshold
pion momenta of 1.4 and 1.7 GeV/c (at which these cross sections possess a high sensitivity to changes in
the in-medium shift $V_0$ of the $K^*(892)^+$ mass)
as well as $\sim$ 15--20\% for incident pion momentum of 2.0 GeV/c,
as our calculations showed. The corrections to the predicted in the paper on the basis of Eq. (17) enhancement
factors (see below) are even smaller. They are about 3--5\% at beam momenta of interest. In view of numerical
results given below, this means that employing of the isotropic distribution (17) in calculations of
the near-threshold $K^*(892)^+$ production in ${\pi^-}A$ reactions with the aim of studying of a possibility
of distinguishing between considered options for the $K^*(892)^+$ in-medium mass shift is very well justified.}$^)$
:
%FORMULA (17), before (14)
\begin{equation}
\frac{d\sigma_{{\pi^{-}}p \to {K^*(892)^+}{\Sigma^-}}(\sqrt{s},<m_{K^*}^*>,m_{\Sigma^-},\theta^*_{K^*})}
{d{\bf \Omega}^*_{K^*}}=\frac{\sigma_{{\pi^{-}}p \to {K^*(892)^+}{\Sigma^-}}(\sqrt{s},\sqrt{s^*_{\rm th}})}{4\pi}.
\end{equation}
Here, $\sigma_{{\pi^{-}}p \to {K^*(892)^+}{\Sigma^-}}(\sqrt{s},\sqrt{s^*_{\rm th}})$ is the
"in-medium" total cross section of channel (1) having the threshold energy $\sqrt{s^*_{\rm th}}$ defined above.
According to the above-mentioned, it is equivalent to the vacuum cross section
$\sigma_{{\pi^{-}}p \to {K^*(892)^+}{\Sigma^-}}(\sqrt{s},\sqrt{s_{\rm th}})$, in which the vacuum threshold energy
$\sqrt{s_{\rm th}}=m_{K^*}+m_{\Sigma^-}=2.089$ GeV is replaced by the in-medium one $\sqrt{s^*_{\rm th}}$
and the free collision energy $s=(E_{\pi^-}+m_N)^2-{\bf p}_{\pi^-}^2$ -- by the in-medium expression (11).
For the free total cross section
$\sigma_{{\pi^{-}}p \to {K^*(892)^+}{\Sigma^-}}(\sqrt{s},\sqrt{s_{\rm th}})$ we have adopted the
following parametrization of the available scarce experimental data [48]:
%formula(18), before (15)
\begin{equation}
\sigma_{{\pi}^-p \to {K^*(892)^+}{\Sigma^-}}(\sqrt{s},\sqrt{s_{\rm th}})=\left\{
\begin{array}{ll}
	67.30\left(\sqrt{s}-\sqrt{s_{\rm th}}\right)^{0.287}~[{\rm {\mu}b}]
	&\mbox{for $0 < \sqrt{s}-\sqrt{s_{\rm th}} \le 0.355~{\rm GeV}$}, \\
	&\\
                   5.66/\left(\sqrt{s}-\sqrt{s_{\rm th}}\right)^{2.103}~[{\rm {\mu}b}]
	&\mbox{for $\sqrt{s}-\sqrt{s_{\rm th}} > 0.355~{\rm GeV}$}.
\end{array}
\right.	
\end{equation}
%%%%%%%%%%%%%%%%%%%%%%%%%%%%%%%%%%%%%%%%%%%%%%%%%%%%%%%%%%%
\begin{figure}[htb]
\begin{center}
\includegraphics[width=12.0cm]{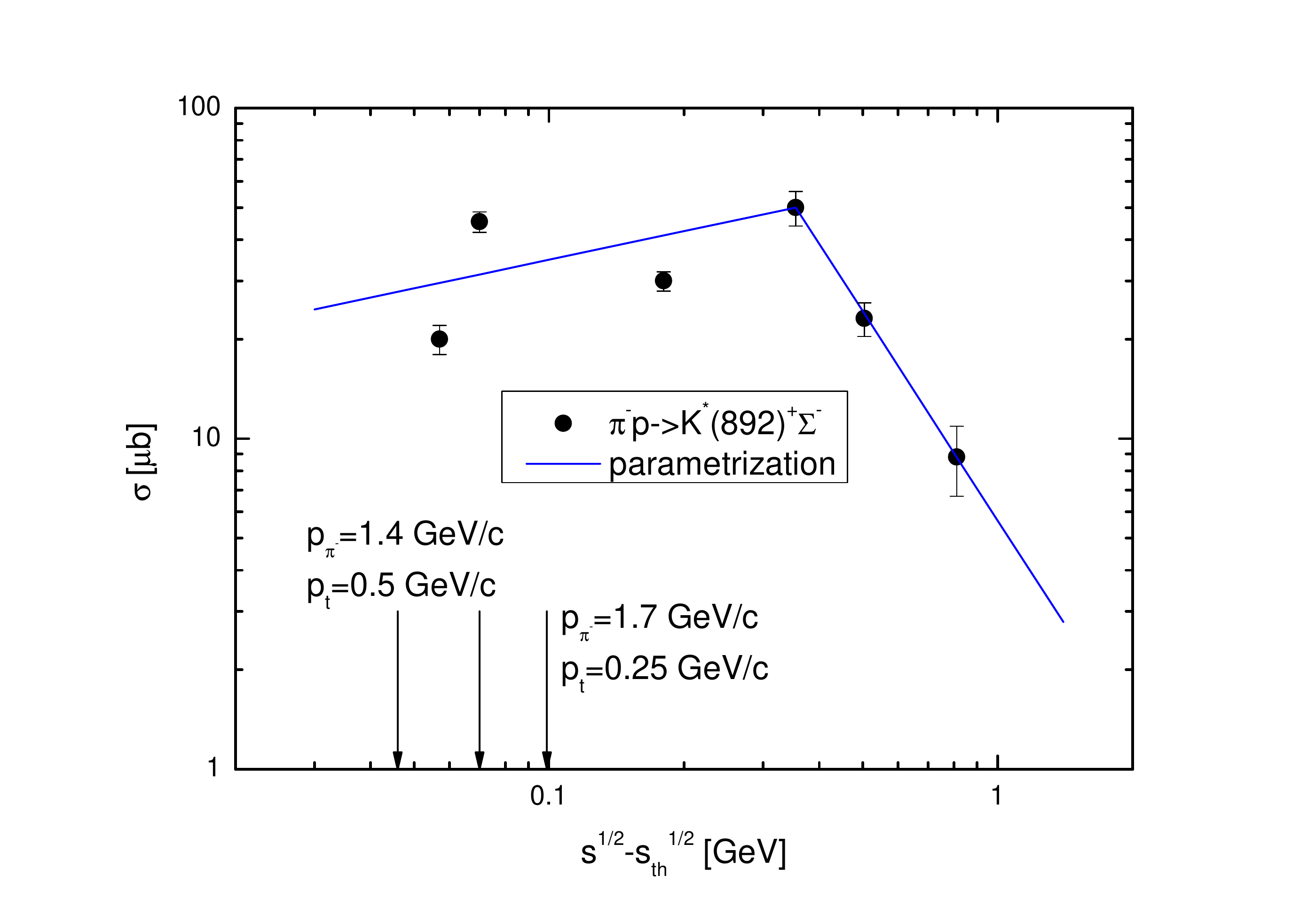}
\vspace*{-2mm} \caption{(color online) Total cross section for the reaction $\pi^-p \to {K^*(892)^+}{\Sigma^-}$
as a function of the excess energy $\sqrt{s}-\sqrt{s_{\rm th}}$. The left and right arrows indicate
the excess energies $\sqrt{s}-\sqrt{s_{\rm th}}$=46 MeV and $\sqrt{s}-\sqrt{s_{\rm th}}$=99 MeV
corresponding to the incident pion momenta of 1.4 and 1.7 GeV/c and a
target proton bound in $^{12}$C by 16 MeV and having momenta of 500 and 250 MeV/c, respectively.
The latter ones are directed opposite to the incoming pion beam. The middle arrow indicates the
excess energy $\sqrt{s}-\sqrt{s_{\rm th}}$=70 MeV corresponding to the initial pion momentum of 2.0 GeV/c
and a free target proton at rest. For the rest of notation see text.}
\label{void}
\end{center}
\end{figure}
%%%%%%%%%%%%%%%%%%%%%%%%%%%%%%%%%%%%%%%%%%%%%%%%%%%%%%%%%%%%%%%%%%%%%%%%%%%%%%%%%%%%%%%%%%%%%
As can be seen from Fig. 1, the parametrization (18) (solid line) fits reasonably well the
data [48] (full circles)
\footnote{$^)$It should be noted that these data correspond to the initial laboratory $\pi^-$ momenta
belonging to the range\\ 1.97 GeV/c $\le p_{\pi^-}\le$ 4.0 GeV/c.}$^)$
for the ${\pi^-}p \to {K^*(892)^+}{\Sigma^-}$ reaction.
One can also see that the on-shell cross section $\sigma_{{\pi}^-p \to {K^*(892)^+}{\Sigma^-}}$
amounts approximately to 31 $\mu$b for the initial pion momentum of 2.0 GeV/c and a free target proton
being at rest. The off-shell cross section $\sigma_{{\pi}^-p \to {K^*(892)^+}{\Sigma^-}}$, calculated
in line with Eqs. (11), (12), (18) for a pion momenta of 1.4 and 1.7 GeV/c and a target proton bound in $^{12}$C
by 16 MeV and having relevant internal momenta of 500 and 250 MeV/c, is about 28 and 35 $\mu$b, respectively
\footnote{$^)$It is interesting to note that the excess energy is equal to -67 MeV for a pion momentum of
1.4 GeV/c and a target proton bound in $^{12}$C by 16 MeV and having internal momentum of 250 MeV/c
directed opposite to the initial pion beam. This means that the main contribution to the deep subthreshold
$K^*(892)^+$ production on nuclei comes from the dynamically formed compact nucleonic configurations -- in
particular, from pairs of correlated $pn$, $pp$ clusters.}$^)$
.
This opens the possibility of measuring the $K^*(892)^+$ yield in $\pi^-A$ reactions both at the
near-threshold and far below threshold beam momenta at the GSI pion beam facility with sizeable strength.
%%%%%%%%%%%%%%%%%%%%%%%%%%%%%%%%%%%%%%%%%%%%%%%%%%%%%%%%%%%%%%%%%%%%%%%%%%%%%%%%%%%%%%%%%%%%%
\begin{figure}[!h]
\begin{center}
\includegraphics[width=16.0cm]{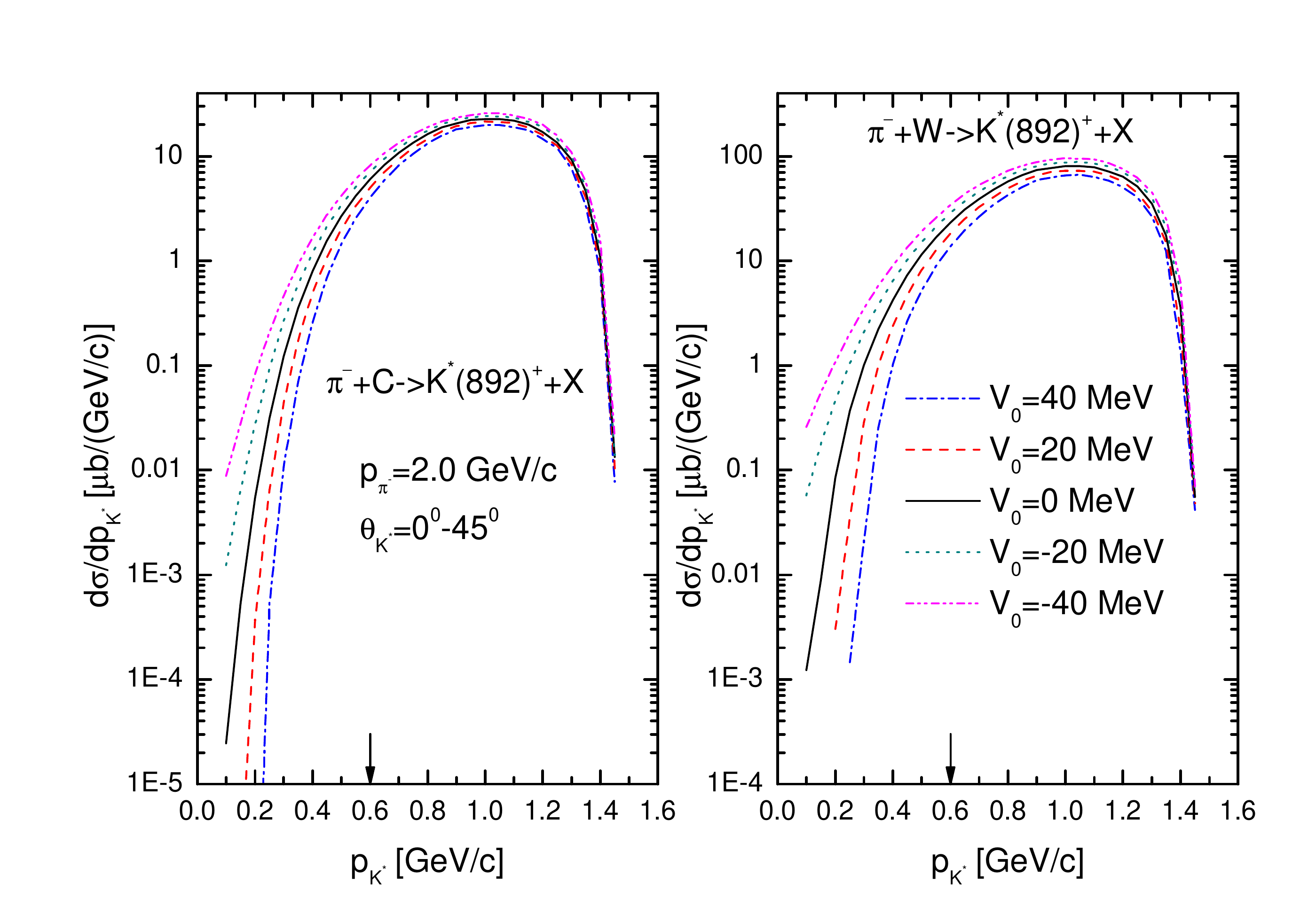}
\vspace*{-2mm} \caption{(color online) Momentum differential cross sections for the production of $K^*(892)^+$
mesons from the primary ${\pi^-}p \to {K^*(892)^+}{\Sigma^-}$ channel in the laboratory polar angular range of
0$^{\circ}$--45$^{\circ}$ in the interaction of $\pi^-$ mesons of momentum of 2.0 GeV/c with $^{12}$C
(left) and $^{184}$W (right) nuclei, calculated for different values of the $K^*(892^+)$ meson effective
scalar potential $V_0$ at density $\rho_0$ indicated in the inset. The arrows indicate the boundary between
the low-momentum and high-momentum regions of the $K^*(892)^+$ spectra.}
\label{void}
\end{center}
\end{figure}
%%%%%%%%%%%%%%%%%%%%%%%%%%%%%%%%%%%%%%%%%%%%%%%%%%%%%%%%%%%
%%%%%%%%%%%%%%%%%%%%%%%%%%%%%%%%%%%%%%%%%%%%%%%%%%%%%%%%%%%
\begin{figure}[!h]
\begin{center}
\includegraphics[width=18.0cm]{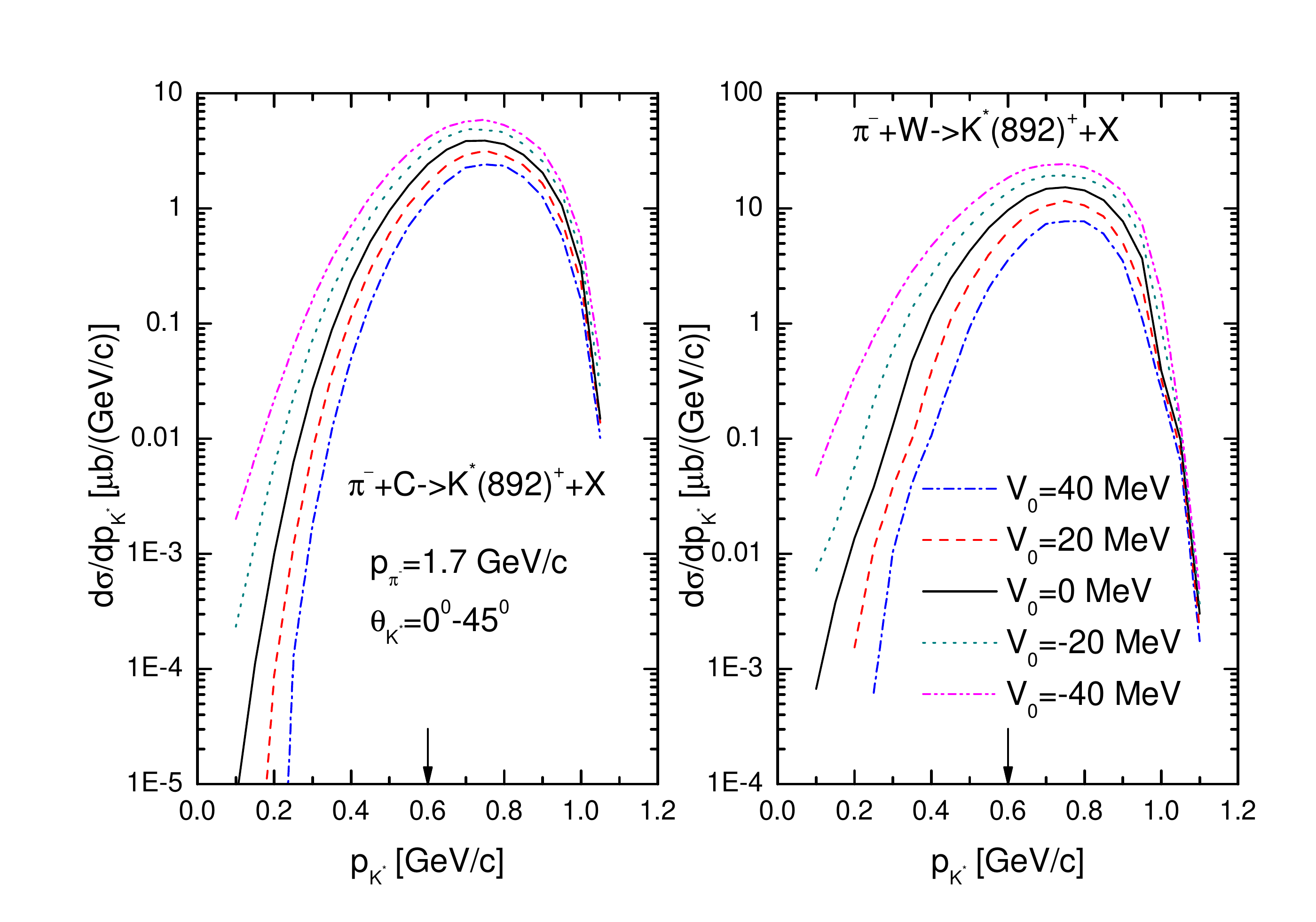}
\vspace*{-2mm} \caption{(color online) The same as in Fig.2, but for the incident pion beam momentum of
1.7 GeV/c.}
\label{void}
\end{center}
\end{figure}
%%%%%%%%%%%%%%%%%%%%%%%%%%%%%%%%%%%%%%%%%%%%%%%%%%%%%%%%%%%%%%%%%%%%%%%%%%%%%%%%%%%%%%%%%%%%%
%%%%%%%%%%%%%%%%%%%%%%%%%%%%%%%%%%%%%%%%%%%%%%%%%%%%%%%%%%%
\begin{figure}[!h]
\begin{center}
\includegraphics[width=18.0cm]{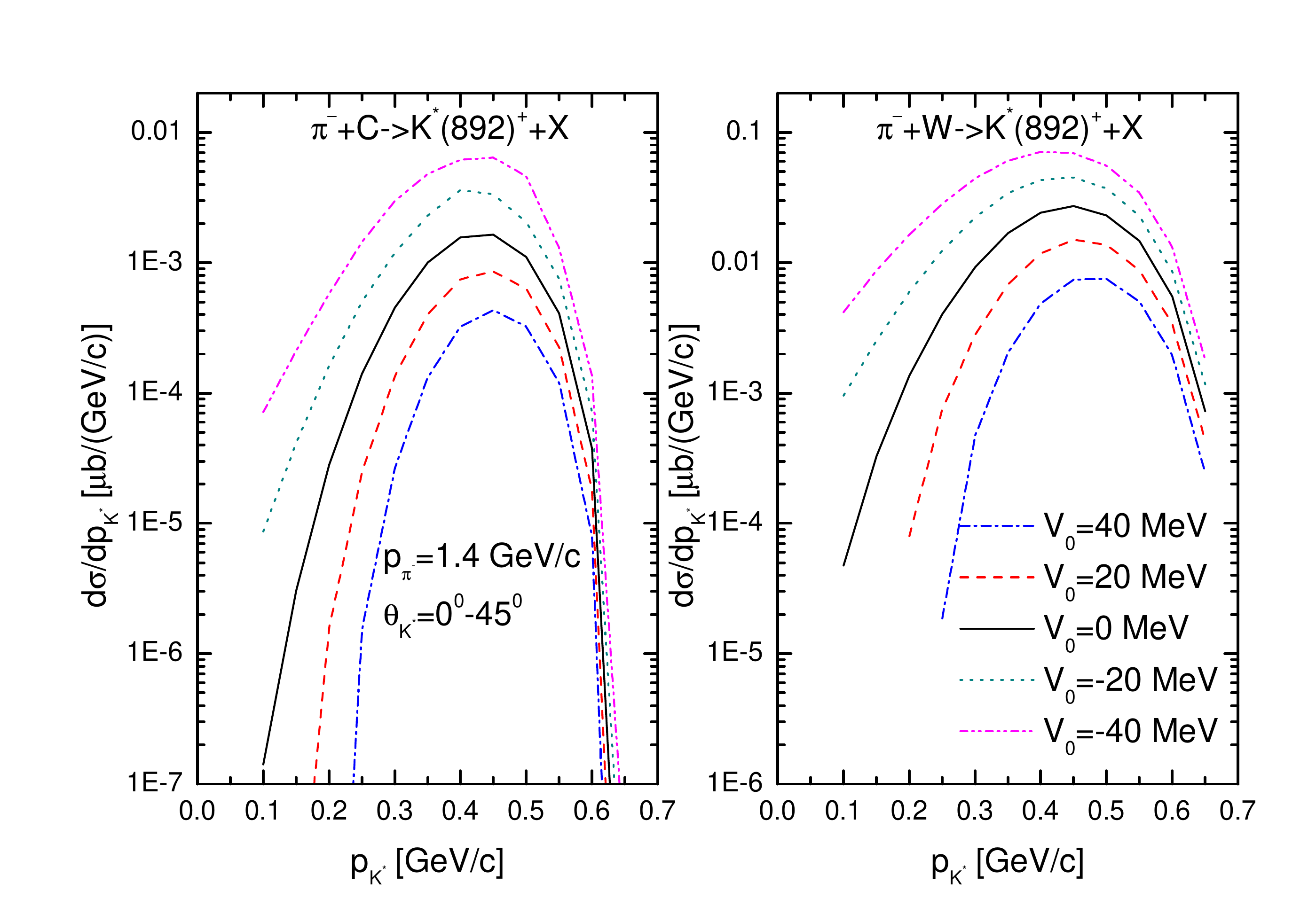}
\vspace*{-2mm} \caption{(color online) The same as in Fig.2, but for the incident pion beam momentum of
1.4 GeV/c.}
\label{void}
\end{center}
\end{figure}
%%%%%%%%%%%%%%%%%%%%%%%%%%%%%%%%%%%%%%%%%%%%%%%%%%%%%%%%%%%%%%%%%%%%%%%%%%%%%%%%%%%%%%%%%%%%%
%%%%%%%%%%%%%%%%%%%%%%%%%%%%%%%%%%%%%%%%%%%%%%%%%%%%%%%%%%%
\begin{figure}[!h]
\begin{center}
\includegraphics[width=18.0cm]{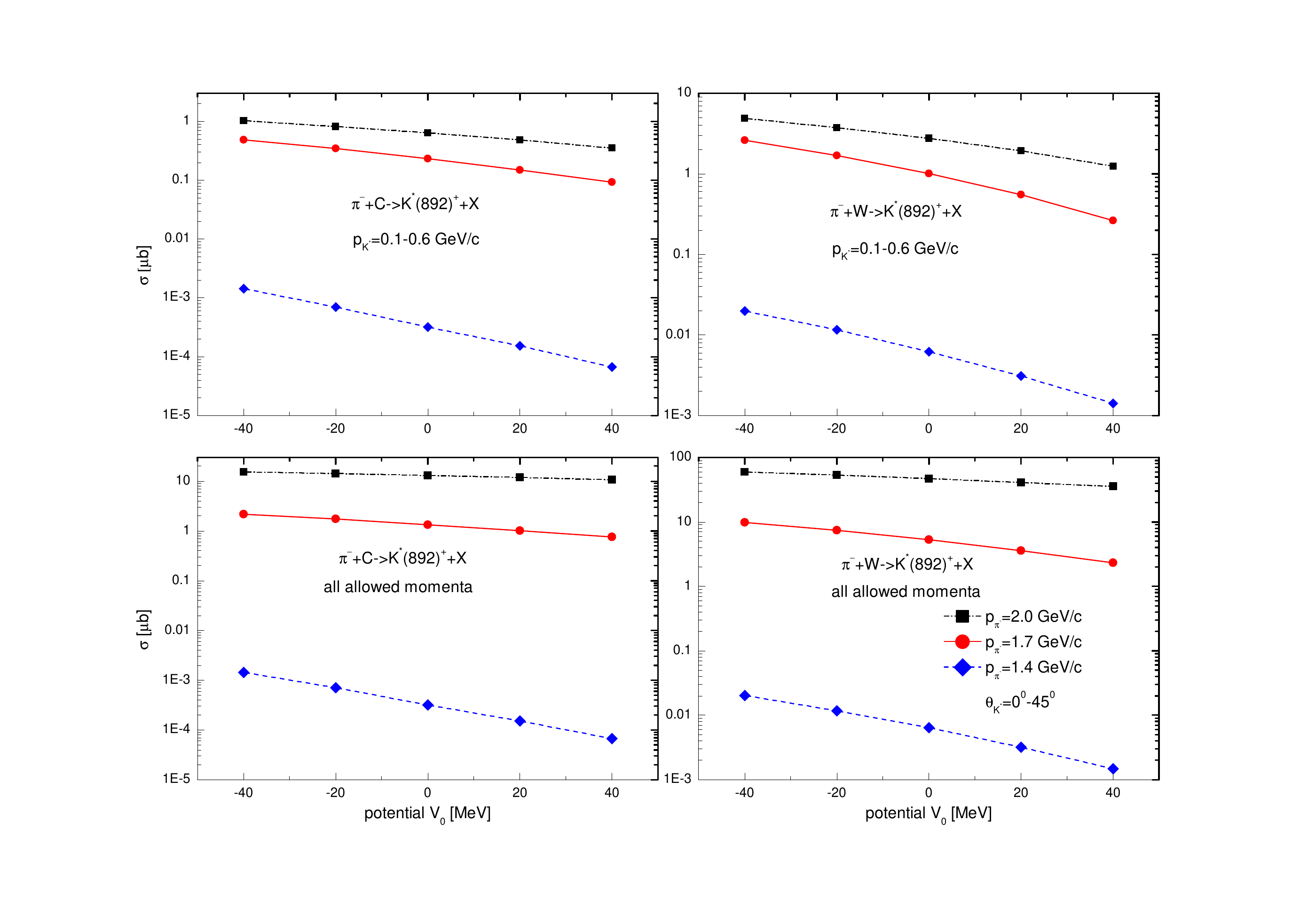}
\vspace*{-2mm} \caption{(color online) The total cross sections for the production of $K^*(892)^+$
mesons from the primary ${\pi^-}p \to {K^*(892)^+}{\Sigma^-}$ channel on C and W target nuclei with
momenta of 0.1--0.6 GeV/c (upper two panels) and with all allowed momenta $\ge$ 0.1 GeV/c at given beam momentum
(lower two panels) in the laboratory polar angular range of 0$^{\circ}$--45$^{\circ}$
by 1.4, 1.7 and 2.0 GeV/c $\pi^-$ mesons as functions of the effective scalar $K^*(892)^+$
potential $V_0$ at normal nuclear density. The lines are to guide the eye.}
\label{void}
\end{center}
\end{figure}
%%%%%%%%%%%%%%%%%%%%%%%%%%%%%%%%%%%%%%%%%%%%%%%%%%%%%%%%%%%%%%%%%%%%%%%%%%%%%%%%%%%%%%%%%%%%%
%%%%%%%%%%%%%%%%%%%%%%%%%%%%%%%%%%%%%%%%%%%%%%%%%%%%%%%%%%%%%%%%%%%%%%%%%%%%%%%%%%%%%%%%%%
\begin{figure}[!h]
\begin{center}
\includegraphics[width=18.0cm]{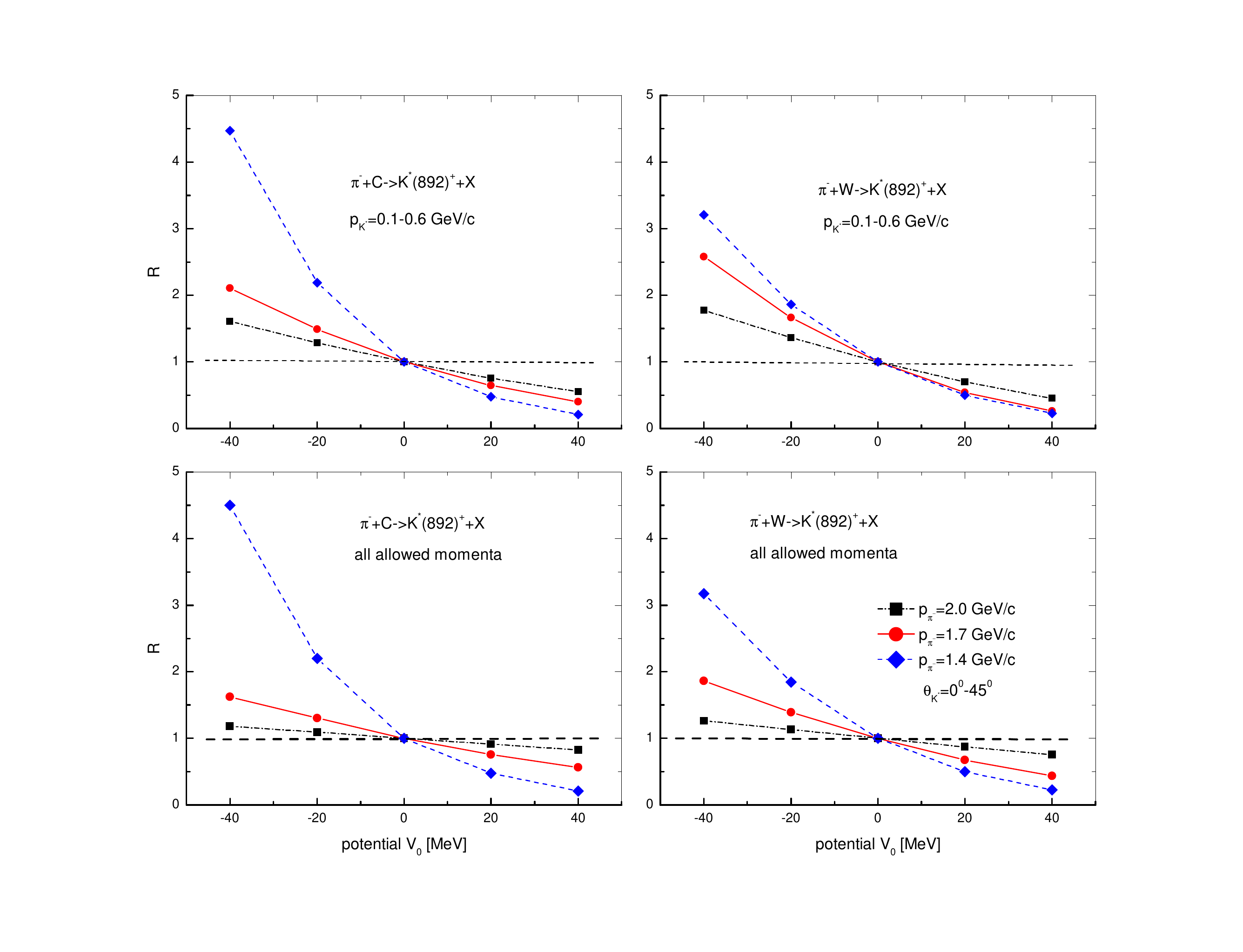}
\vspace*{-2mm} \caption{(color online) Ratio between the total cross sections for the production of $K^*(892)^+$
mesons from the primary ${\pi^-}p \to {K^*(892)^+}{\Sigma^-}$ channel
on $^{12}$C and $^{184}$W target nuclei at laboratory angles of 0$^{\circ}$--45$^{\circ}$ with
momenta of 0.1--0.6 GeV/c (upper two panels) and with all allowed momenta $\ge$ 0.1 GeV/c at given beam momentum
(lower two panels) by 1.4, 1.7 and 2.0 GeV/c $\pi^-$ mesons, calculated with and without the $K^*(892)^+$
in-medium mass shift $V_0$ at normal nuclear density, as function of this shift. The lines are to guide the eye.}
\label{void}
\end{center}
\end{figure}
%%%%%%%%%%%%%%%%%%%%%%%%%%%%%%%%%%%%%%%%%%%%%%%%%%%%%%%%%%%%%%%%%%%%%%%%%%%%%%%%%%%%%%%%%%%%%
%%%%%%%%%%%%%%%%%%%%%%%%%%%%%%%%%%%%%%%%%%%%%%%%%%%%%%%%%%%%%%%%%%%%%%%%%%%%%%%%%%%%%%%%%%%%%
\begin{figure}[!h]
\begin{center}
\includegraphics[width=18.0cm]{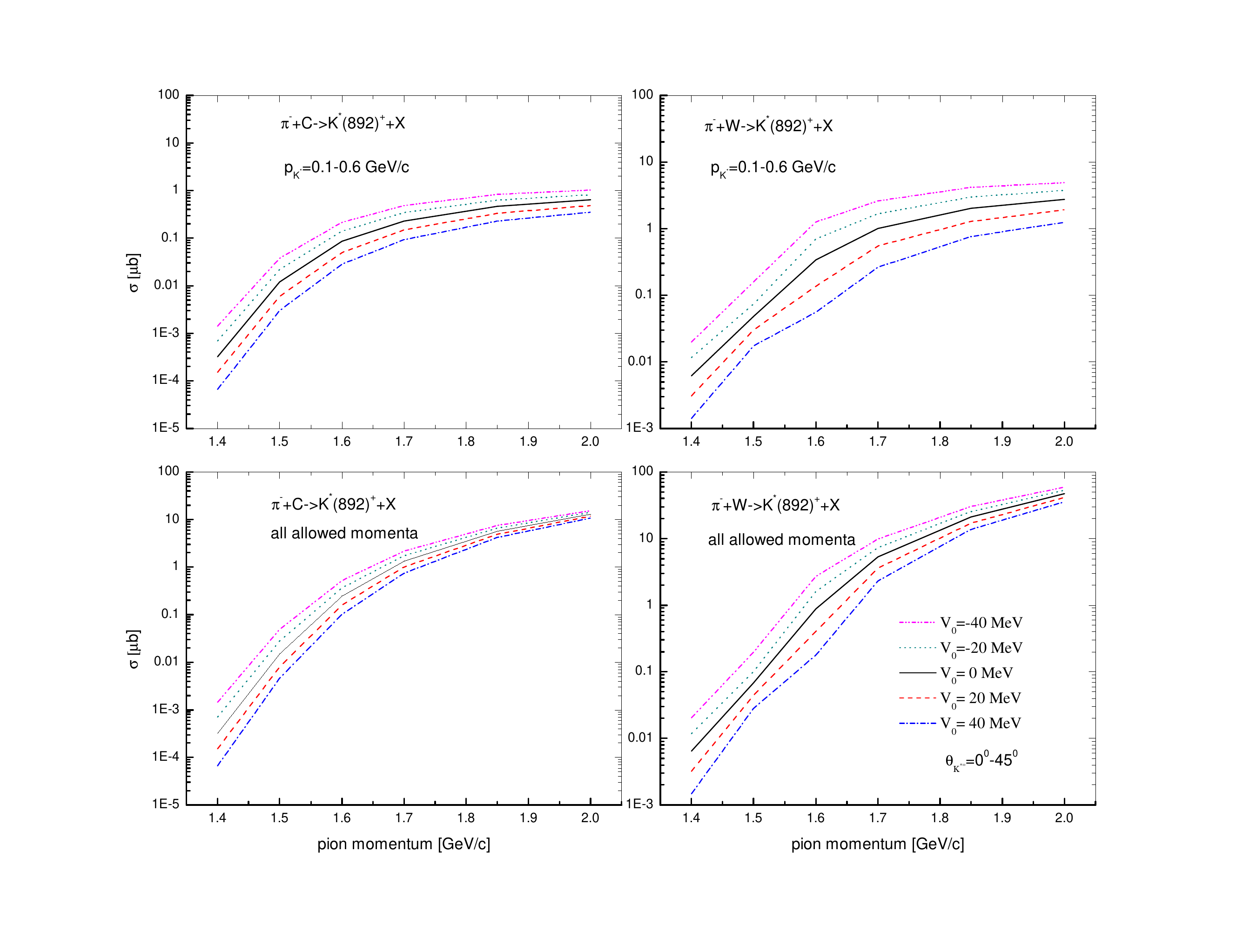}
\vspace*{-2mm} \caption{(color online) The total cross sections for the production of $K^*(892)^+$
mesons from the primary ${\pi^-}p \to {K^*(892)^+}{\Sigma^-}$ channel
on $^{12}$C and $^{184}$W target nuclei at laboratory angles of 0$^{\circ}$--45$^{\circ}$ with
momenta of 0.1--0.6 GeV/c (upper two panels) and with all allowed momenta $\ge$ 0.1 GeV/c at given beam momentum
(lower two panels), calculated with the $K^*(892)^+$ in-medium mass shift $V_0$ at normal nuclear density depicted
in the inset, as functions of the incident pion momentum.}
\label{void}
\end{center}
\end{figure}
%%%%%%%%%%%%%%%%%%%%%%%%%%%%%%%%%%%%%%%%%%%%%%%%%%%%%%%%%%%%%%%%%%%%%%%%%%%%%%%%%%%%%%%%%%
In Eqs. (6)--(8) we assume that the direction of the $K^*(892)^+$ three-momentum is not changed
during the propagation from its production point inside the nucleus  in the relatively weak nuclear
field, considered in the work, to the vacuum far away from the nucleus. As a consequence, the
quantities
$\left<d\sigma_{{\pi^-}p\to K^*(892)^+{{\Sigma^-}}}({\bf p}_{\pi^-},
{\bf p}^{\prime}_{{K^*}})/d{\bf p}^{\prime}_{{K^*}}\right>_A$ and
$d{\bf p}^{\prime}_{{K^*}}/d{\bf p}_{{K^*}}$, entering into Eq. (5), can be put in the
simple forms
$\left<d\sigma_{{\pi^-}p\to K^*(892)^+{{\Sigma^-}}}(p_{\pi^-},
p^{\prime}_{{K^*}}, \theta_{K^*})/p^{\prime2}_{K^*}dp^{\prime}_{{K^*}}d{\bf \Omega}_{K^*}\right>_A$ and
$p^{\prime}_{{K^*}}/p_{{K^*}}$, where
${\bf \Omega}_{K^*}(\theta_{K^*},\varphi_{K^*})={\bf p}_{{K^*}}/p_{{K^*}}$.
Here, $\varphi_{K^*}$ is the azimuthal angle of the $K^*(892)^+$ momentum ${\bf p}_{K^*}$
in the laboratory system.
Accounting for the HADES spectrometer acceptance as well as the fact that in the considered energy
region $K^*(892)^+$ mesons are mainly emitted, due to the kinematics, in forward directions
\footnote{$^)$ Thus, for instance, at a beam momentum of 2.0 GeV/c the $K^*(892)^+$ laboratory production
polar angles in reaction (1) proceeding on the target proton being at rest are $\le$ 19$^{\circ}$.}$^)$
,
we will calculate the $K^*(892)^+$ momentum
differential and total production cross sections on $^{12}$C and $^{184}$W target nuclei
for laboratory solid angle
${\Delta}{\bf \Omega}_{K^*}$=$0^{\circ} \le \theta_{K^*} \le 45^{\circ}$,
and $0 \le \varphi_{K^*} \le 2{\pi}$.
Integrating the full inclusive differential cross section (5) over this angular domain,
we can represent the differential cross section for $K^*(892)^+$ meson
production in ${\pi^-}A$ collisions from the direct process (1), corresponding to the
HADES acceptance window
\footnote{$^)$ At HADES the $K^*(892)^+$ mesons could be identified via the hadronic decays
$K^*(892)^+ \to K^0{\pi^+}$ with a branching ratio of 2/3 or via their radiative decays
$K^*(892)^+ \to K^+{\gamma}$ with sizable branching ratio of 10$^{-3}$ [49].}$^)$
, in the following form:
%formula(19), before (16)
\begin{equation}
\frac{d\sigma_{{\pi^-}A\to {K^*(892)^+}X}^{({\rm prim})}
(p_{\pi^-},p_{K^*})}{dp_{K^*}}=
\int\limits_{{\Delta}{\bf \Omega}_{K^*}}^{}d{\bf \Omega}_{K^*}
\frac{d\sigma_{{\pi^-}A\to {K^*(892)^+}X}^{({\rm prim})}
({\bf p}_{\pi^-},{\bf p}_{K^*})}{d{\bf p}_{K^*}}p_{K^*}^2
\end{equation}
$$
=2{\pi}\left(\frac{Z}{A}\right)\left(\frac{p_{K^*}}{p^{\prime}_{K^*}}\right)
\int\limits_{\cos45^{\circ}}^{1}d\cos{{\theta_{K^*}}}I_{V}[A,\theta_{K^*}]
\left<\frac{d\sigma_{{\pi^-}p\to {K^*(892)^+}{\Sigma^-}}(p_{\pi^-},
p^{\prime}_{K^*},\theta_{K^*})}{dp^{\prime}_{K^*}d{\bf \Omega}_{K^*}}\right>_A.
$$

\section*{3 Numerical results and discussion}

\hspace{0.5cm} In the beginning, we consider the absolute $K^*(892)^+$ momentum differential
cross sections from the direct $K^*(892)^+$ production mechanism
in $\pi^-$$^{12}$C and $\pi^-$$^{184}$W collisions.
These cross sections were calculated according to Eq. (19) in five
considered scenarios for the $K^*(892)^+$ in-medium mass shift at density $\rho_0$ at laboratory angles
of 0$^{\circ}$--45$^{\circ}$ and for incident pion momenta of 2.0, 1.7 and 1.4 GeV/c. They are presented,
respectively, in Figs. 2, 3 and 4. It is seen that the $K^*(892)^+$ meson momentum distributions are appreciable
sensitive to its in-medium mass shift mainly in the low-momentum region of 0.1--0.6 GeV/c for both target nuclei
and for all considered beam momenta. Here there are a sizeable and experimentally accessible differences
between the results obtained by employing different $K^*(892)^+$ in-medium mass shifts under consideration,
which for each target nucleus are practically similar to each other at these initial pion momenta.
Thus, for example, for incident pion and outgoing $K^*(892)^+$ meson momenta of 2.0 and 0.3 GeV/c, respectively,
and in the case of $^{12}$C nucleus the $K^*(892)^+$ yield is enhanced at mass shift $V_0=+20$ MeV by about
a factor of 4.1 as compared to that obtained for the shift $V_0=+40$ MeV. When going from $V_0=+20$ MeV to
$V_0=0$ MeV, from $V_0=0$ MeV to $V_0=-20$ MeV and from $V_0=-20$ MeV to $V_0=-40$ MeV the enhancement factors are
about 2.8, 2.2 and 1.8. In the case of $^{184}$W target nucleus these enhancement factors are about 14.0, 3.5, 2.1
and 1.7. At initial beam momentum of 1.4 GeV/c and the same outgoing kaon momentum of 0.3 GeV/c the corresponding
enhancement factors are similar and are about 5.0, 3.3, 2.6 and 2.5 as well as 6.0, 3.3, 2.4 and 2.0 in the cases
of $^{12}$C as well as $^{184}$W target nuclei, respectively. However, the $K^*(892)^+$ low-momentum production
differential cross sections at beam momentum of 1.4 GeV/c are very small
(in the range of $\sim$ 0.0001--0.1 ${\rm \mu}$b/(GeV/c)) and they are less than those at pion momenta of 1.7 and
2.0 GeV/c by about of two--three orders of magnitude. Therefore, the measurements of the $K^*(892)^+$ differential
cross sections in ${\pi^-}A$ reactions in the near-threshold incident pion momentum region (at 1.7--2.0 GeV/c) with
the aim of distinguishing between considered options for the $K^*(892)^+$ mass shift in nuclear matter look promising.

The sensitivity of the low-momentum parts of the $K^*(892)^+$ meson production differential cross sections to
its in-medium mass shift $V_0$, shown in Figs. 2, 3, 4, can be also studied from such integral measurements
as the measurements of the total cross sections for $K^*(892)^+$ production in ${\pi^-}^{12}$C
and  ${\pi^-}^{184}$W reactions by 1.4, 1.7 and 2.0 GeV/c pions at laboratory angles of 0$^{\circ}$--45$^{\circ}$
in the low-momentum (0.1--0.6 GeV/c) and in the allowed for given beam momentum full-momentum regions.
These cross sections, calculated by integrating Eq. (19) over the $K^*(892)^+$ momentum $p_{K^*}$
in these regions, are shown in Fig. 5 as functions of the mass shift (or effective scalar potential) $V_0$.
It can be seen from this figure that again the low-momentum range of 0.1--0.6 GeV/c shows the highest sensitivity
to this potential. Thus, for instance, the ratios between the total cross sections of $K^*(892)^+$ production by
1.4, 1.7, 2.0 GeV/c pions on $^{12}$C and $^{184}$W target nuclei in this momentum range, calculated with the
potential $V_0=-40$ MeV, and the same cross sections, obtained in the scenario $V_0=+40$ MeV, respectively, are
about 21.0, 5.0, 3.0 and 14.0, 10.0, 4.0. While the same ratios in the full-momentum regions are about
21.0, 3.0, 1.4 for $^{12}$C and 14.0, 4.0, 1.7 for $^{184}$W. In the low-momentum region of interest the highest
sensitivity of the $K^*(892)^+$ production total cross sections to the potential $V_0$ is observed, as is expected,
at initial pion momentum of 1.4 GeV/c. However, these cross sections are small and they are less than those at
beam momenta of 1.7 and 2.0 GeV/c by several orders of magnitude. Since the latter ones have a measurable strengh
$\sim$ 0.1--5 ${\rm \mu}$b, the low-momentum total cross section measurements of $K^*(892)^+$ meson production on
nuclei in the near-threshold incident pion momentum region $\sim$ 1.7--2.0 GeV/c with the aim of distinguishing
between adopted options for its mass shift in nuclear matter look promising as well.

The fact that the low-momentum range of 0.1--0.6 GeV/c shows the highest sensitivity to the $K^*(892)^+$
in-medium mass shift $V_0$ at the central density $\rho_0$ is clearly supported also by the results given
in Fig. 6. Here, the ratios $R$ of the $K^*(892)^+$ meson production total cross sections calculated for its
mass shift $V_0$ and presented in Fig. 5 to the analogous cross sections determined at $V_0=0$ MeV are shown
as functions of this mass shift. It is worth noting that an analysis of these ratios has the advantage that
they do not depend on the absolute normalization of calculated and measured cross sections.
As is seen from this figure, the highest sensitivity of the ratios in both considered kinematic ranges to the
quantity $V_0$ is indeed observed at pion momentum of 1.4 GeV/c. For example, at this momentum and for these
ranges the cross section ratios R for $V_0=-40$ MeV are about 4.5 and 3.2 for $^{12}$C and $^{184}$W, respectively.
As the pion-beam momentum increases to 1.7 and 2.0 GeV/c, the sensitivity of the cross section ratios to
variations in the mass shift $V_0$ becomes somewhat lower. Thus, in the case where $K^*(892)^+$ mesons of
momenta of 0.1--0.6 GeV/c are produced by 1.7 and 2.0 GeV/c pions incident to a $^{12}$C as well as $^{184}$W
targets, the ratios being considered take for $V_0=-40$ MeV smaller but yet a sizeable values of 2.1
and 1.6 as well as 2.6 and 1.8, respectively. The analogous ratios for the production of the $K^*(892)^+$
mesons in the full-momentum regions by 1.7 and 2.0 GeV/c pions on $^{12}$C as well as $^{184}$W nuclei are
somewhat yet smaller. Namely, they are about 1.6 and 1.2 as well as 1.9 and 1.3, respectively.

Therefore, we come to the conclusion that a comparison of the low-momentum "integral" results shown in
Figs. 5, 6 with the respective near-threshold experimental data, which could be taken in future experiments
using $\pi^-$ beams at the GSI pion beam facility or at J-PARC [50],
will also allow to study the in-medium properties of the $K^*(892)^+$ mesons.

These properties can be also investigated [5] from such another integral meaurements as the measurements
of the excitation functions for $K^*(892)^+$ production in ${\pi^-}^{12}$C and ${\pi^-}^{184}$W reactions
at laboratory angles of 0$^{\circ}$--45$^{\circ}$ in the low-momentum (0.1--0.6 GeV/c) and in the
full-momentum regions. They were calculated for five adopted scenarios for the $K^*(892)^+$ in-medium
mass shift and are given in Fig. 7. One can see that the absolute values of the excitation functions show
a wider variation for the mass shift range of $V_0=-40$ to +40 MeV in the low-momentum region for all
considered beam momenta. In this momentum region and at beam momenta not far below the threshold
(at $p_{\pi^-}$ $\sim$ 1.6--1.84 GeV/c) there are well separated and experimentally distinguishable
differences ($\sim$ 25--45\% for $^{12}$C and $\sim$ 30--60\% for $^{184}$W) between all calculations
corresponding to different options for the $K^*(892)^+$ in-medium mass shift. Here, the total $K^*(892)^+$
production cross sections have a measurable strength $\sim$ 30--3000 nb. At above threshold pion momenta
of 1.84--2.0 GeV/c the impact of the $K^*(892)^+$ meson mass shift on its yield becomes somewhat lower.
Here, the respective differences are $\sim$ 20--30\% for $^{12}$C and $\sim$ 25--35\% for $^{184}$W,
but one might expect to measure their as well in future experiments at the GSI pion beam facility
\footnote{$^)$Since, as one may hope, the precision of these experiments can reach the same value $\sim$
15\% as was achieved in recent measurements here [39] of $\pi^-$ meson-induced
$K^+$ meson production in ${\pi^-}^{12}C \to K^+X$ and ${\pi^-}^{184}W \to K^+X$ reactions at 1.7 GeV/c
beam momentum.}$^)$
.

   Taking into account the above consideration, one can conclude that the near-threshold
$K^*(892)^+$ differential and total cross section measurements at $K^*(892)^+$ momenta of 0.1--0.6 GeV/c
in ${\pi^-}A$ interactions
will allow to shed light on the possible $K^*(892)^+$ in-medium mass shift at these momenta.

\section*{4 Conclusions}

\hspace{0.5cm} In this paper we study the inclusive strange vector meson $K^*(892)^+$ production
in ${\pi^-}A$ reactions at near-threshold laboratory incident pion momenta of 1.4--2.0 GeV/c
within a nuclear spectral function approach.
The approach accounts for incoherent primary $\pi^-$ meson--proton
${\pi^-}p \to {K^*(892)^+}\Sigma^-$ production processes as well as the influence of the scalar
$K^*(892)^+$--nucleus potential (or the $K^*(892)^+$ in-medium mass shift) on these processes.
We calculate the absolute differential and total cross
sections for the production of $K^*(892)^+$ mesons on carbon and tungsten target nuclei at laboratory angles of
0$^{\circ}$--45$^{\circ}$ and at these initial pion momenta
within five scenarios for the above shift. We show that the $K^*(892)^+$ momentum distributions and their
excitation functions (absolute and relative) possess a high sensitivity to changes in the in-medium
$K^*(892)^+$ mass shift in the low-momentum region of 0.1--0.6 GeV/c.
Therefore, the measurement of such observables in a dedicated experiment at
the GSI pion beam facility in the near-threshold momentum domain will allow to get valuable information
on the $K^*(892)^+$ in-medium properties.
\\
\\

%%%%%%%%%%%%%%%%%%%%%%%%%%%%%%%%%%%%%%%%%%%%%%%%%%%%%%%%%%%%%%%%

\begin{thebibliography}{99}
\bibitem{1} R. Rapp and J. Wambach, Adv. Nucl. Phys. {\bf 25}, 1 (2000);\\
                            arXiv:hep-ph/9909229.
\bibitem{2} R. S. Hayano and T. Hatsuda, Rev. Mod. Phys. {\bf 82}, 2949 (2010);\\
                            arXiv:0812.1702 [nucl-ex].
\bibitem{3} S. Leupold, V. Metag, and U. Mosel, Int. J. Mod. Phys. E {\bf 19}, 147 (2010);\\
                            arXiv:0907.2388 [nucl-th].
\bibitem{4} G. Krein, A. W. Thomas, and K. Tsushima, Prog. Part. Nucl. Phys. {\bf 100}, 161 (2018);\\
                            arXiv:1706.02688 [hep-ph].
\bibitem{5} V. Metag, M. Nanova, and E. Ya. Paryev, Prog. Part. Nucl. Phys. {\bf 97}, 199 (2017);\\
                            arXiv:1706.09654 [nucl-ex].
\bibitem{6} A. Gal, E. V. Hungerford and D. J. Millener, Rev. Mod. Phys. {\bf 88}, 035004 (2016);\\
                            arXiv:1605.00557 [nucl-th].
\bibitem{7} K. Tsushima {\it et al.}, Phys. Rev. C {\bf 83}, 065208 (2011) [arXiv:1103.5516 [nucl-th]];\\
            G. Krein, A. W. Thomas, and K. Tsushima, Phys. Lett. B {\bf 697}, 136 (2011)
                            [arXiv:1007.2220 [nucl-th]].
\bibitem{8} E. Ya. Paryev, Yu. T. Kiselev, and Yu. M. Zaitsev, Nucl. Phys. A {\bf 968}, 1 (2017);\\
            E. Ya. Paryev and Yu. T. Kiselev, Nucl. Phys. A {\bf 978}, 201 (2018) [arXiv:1810.01715 [nucl-th]];\\
            E. Ya. Paryev and Yu. T. Kiselev, Phys. Atom. Nucl. Vol. {\bf 80}, No.1, 67 (2017);\\
            E. Ya. Paryev and Yu. T. Kiselev, Phys. Atom. Nucl. Vol. {\bf 81}, No.5, 566 (2018);\\
            E. Ya. Paryev, Nucl. Phys. A {\bf 996}, 121711 (2020) [arXiv:2003.00788 [nucl-th]].
\bibitem{9} E. Ya. Paryev, Chinese Physics C, Vol. {\bf 42}, No. (8), 084101 (2018);\\
                             arXiv:1806.00303 [nucl-th].
\bibitem{10} E. Ya. Paryev, Nucl. Phys. A {\bf 988}, 24 (2019);\\
                            arXiv:1906.02185 [nucl-th].
\bibitem{11} S. D. Bass and P. Moskal, arXiv:1810.12290 [hep-ph].
\bibitem{12} M. Nanova {\it et al.}, Eur. Phys. J. A {\bf 54}: 182 (2018);\\
                            arXiv:1810.01288 [nucl-ex].
\bibitem{13} N. Tomida {\it et al.}, Phys. Rev. Lett. {\bf 124}, 202501 (2020);\\
                            arXiv:2005.03449 [nucl-ex].
\bibitem{14} M. M. Kaskulov and E. Oset, Phys. Rev. C {\bf 73}, 045213 (2006);\\
                            arXiv:nucl-th/0509088.
\bibitem{15} M. M. Kaskulov and E. Oset, AIP Conf. Proc. {\bf 842}, 483--5 (2006).
\bibitem{16} M. F. M. Lutz, C. L. Copra and M. Moeller, Nucl. Phys. A {\bf 808}, 124 (2008);\\
                            arXiv:0707.1283 [nucl-th].
\bibitem{17} D. Cabrera {\it et al.}, Phys. Rev. C {\bf 90}, 055207 (2014);\\
                            arXiv:1406.2570 [hep-ph].
\bibitem{18} S. Petschauer {\it et al.}, Eur. Phys. J. A {\bf 52}, 15 (2016);\\
                            arXiv:1507.08808 [nucl-th].
\bibitem{19} E. Ya. Paryev, M. Hartmann, and Yu. T. Kiselev, Chinese Physics C,\\
                             Vol. {\bf 41}, No. (12), 124108 (2017);\\
                             arXiv:1612.02767 [nucl-th].
\bibitem{20} Z. Q. Feng, W. J. Xie, and G. M. Jin, Phys. Rev. C {\bf 90}, 064604 (2014).
\bibitem{21} E. Ya. Paryev and Yu. T. Kiselev, Nucl. Phys. A {\bf 992}, 121622 (2019);\\
                            arXiv:1910.02755 [nucl-th].
\bibitem{22} M. Kaskulov, L. Roca and E. Oset, Eur. Phys. J. A {\bf 28}, 139 (2006);\\
                            arXiv:nucl-th/0601074.
\bibitem{23} E. Ya. Paryev, Phys. Atom. Nucl. Vol. {\bf 75}, No.12, 1523 (2012).
\bibitem{24} E. Ya. Paryev, J. Phys. G: Nucl. Part. Phys. {\bf 37}, 105101 (2010);\\
                            arXiv:1010.0111 [nucl-th].
\bibitem{25} S. H. Lee and S. Cho, Int. J. Mod. Phys. E {\bf 22}, 1330008 (2013);\\
                            arXiv:1302.0642 [nucl-th].
\bibitem{26} T. Song, T. Hatsuda, and S. H. Lee, Phys. Lett. B {\bf 792}, 160 (2019);\\
                            arXiv:1808.05372 [nucl-th].
\bibitem{27} S. H. Lee, arXiv:1904.09064 [nucl-th].
\bibitem{28} L. Tolos, R. Molina, E. Oset, and A. Ramos, Phys. Rev. C {\bf 82}, 045210 (2010);\\
                                        arXiv:1006.3454 [nucl-th].
\bibitem{29} E. Oset {\it et al.}, Int. J. Mod. Phys. E {\bf 21}, 1230011 (2012);\\
                                        arXiv:1210.3738 [nucl-th].
\bibitem{30} A. Ilner, D. Cabrera, P. Srisawad, and E. Bratkovskaya, Nucl. Phys. A {\bf 927}, 249 (2014);\\
                                        arXiv:1312.5215 [hep-ph].
\bibitem{31} D. Cabrera {\it et al.}, Journal of Physics: Conf. Series {\bf 503}, 012017 (2014);\\
                                        arXiv:1312.4343 [hep-ph].\\
             L. Tolos, EPJ Web of Conf. {\bf 171}, 09003 (2018).
\bibitem{32} K. Tsushima, A. Sibirtsev, and A. W. Thomas, Phys. Rev. C {\bf 62}, 064904 (2000);\\
                                        arXiv:nucl-th/0004011.
\bibitem{33} D. Suenaga and P. Lakaschus, Phys. Rev. C {\bf 101}, 035209 (2020);\\
                                        arXiv:1908.10509 [nucl-th].
\bibitem{34} T. Anticic {\it et al.} (NA 49 Collaboration), Phys. Rev. C {\bf 84}, 064909 (2011);\\
                                        arXiv:1105.3109 [nucl-ex].
\bibitem{35} J. Adams {\it et al.} (STAR Collaboration), Phys. Rev. C {\bf 71}, 064902 (2005)
                                      [arXiv:nucl-ex/0412019];\\
             M. M. Aggarwal {\it et al.} (STAR Collaboration), Phys. Rev. C {\bf 84}, 034909 (2011)
                                      [arXiv:1006.1961 [nucl-ex]].
\bibitem{36} B. Abelev {\it et al.} (ALICE Collaboration), Eur. Phys. J. C {\bf 72}, 2183 (2012).
\bibitem{37} X. Lopez {\it et al.} (FOPI Collaboration), Phys. Rev. C {\bf 81}, 061902 (2010);\\
                                    arXiv:1006.1905 [nucl-ex].
\bibitem{38} G. Agakishiev {\it et al.} (HADES Collaboration), Eur. Phys. J. A {\bf 49}: 34 (2013).
\bibitem{39} J. Adamczewski-Musch {\it et al.} (HADES Collaboration), Phys. Rev. Lett. {\bf 123}, 022002 (2019);\\
                                    arXiv:1812.03728 [nucl-ex].
\bibitem{40} J. Haidenbauer and Ulf-G Meissner, Nucl. Phys. A {\bf 936}, 29 (2015);\\
                                    arXiv:1411.3114 [nucl-th].
\bibitem{41} M. Kohno and Y. Fujiwara, Phys. Rev. C {\bf 79}, 054318 (2009);\\
                                    arXiv:0904.0517 [nucl-th].
\bibitem{42} M. Kohno, Phys. Rev. C {\bf 81}, 014003 (2010);\\
                                    arXiv:0912.4330 [nucl-th].
\bibitem{43} E. Friedman and A. Gal, Phys. Rep. {\bf 452}, 89 (2007);\\
                                    arXiv:0705.3965 [nucl-th].
\bibitem{44} K. P. Khemchandani {\it et al.}, Phys. Rev. D {\bf 91}, 094008 (2015);\\
                                    arXiv:1406.7203 [nucl-th].
\bibitem{45} S. V. Efremov and  E. Ya. Paryev, Eur. Phys. J. A {\bf 1}, 99 (1998).
\bibitem{46} E. Ya. Paryev, Eur. Phys. J. A {\bf 9}, 521 (2000).
\bibitem{47} E. Ya. Paryev, Eur. Phys. J. A {\bf 7}, 127 (2000).
\bibitem{48} V. Flaminio {\it et al.}, Compilation of Cross Sections.\\
             I: ${\pi^+}$ and ${\pi^-}$ Induced Reactions. CERN-HERA {\bf 83-01}, (1983).
\bibitem{49} T. Hatsuda, arXiv:nucl-th/9702002.
\bibitem{50} M. Moritsu {\it et al.} (J-PARC E19 Collaboration), Phys. Rev. C {\bf 90}, 035205 (2014);\\
                                    arXiv:1407.0669 [nucl-ex].
\end{thebibliography}
\end{document}